# An Overview of Automated Vehicle Longitudinal Platoon Formation Strategies[§]

M Sabbir Salek, M.S., Salek[*]

Senior Engineer, USDOT National Center for Transportation Cybersecurity and Resiliency, Greenville, SC 29607, USA, msalek@clemson.edu

Mugdha Basu Thakur, M.B., Thakur

Ph.D. Candidate, Department of Automotive Engineering, Clemson University, Greenville, SC 29607, USA, mbasuth@clemson.edu

Pardha Sai Krishna Ala, P.S.K., Ala

Ph.D. Candidate, Department of Automotive Engineering, Clemson University, Greenville, SC 29607, USA, pala@clemson.edu

Mashrur Chowdhury, M., Chowdhury

Professor, Glenn Department of Civil Engineering, Clemson University, Clemson, SC 29634, USA, mac@clemson.edu

Matthias Schmid, M., Schmid

Assistant Professor, Department of Automotive Engineering, Clemson University, Greenville, SC 29607, USA, schmidm@clemson.edu

Pamela Murray-Tuite, P., Murray-Tuite

Professor, Glenn Department of Civil Engineering, Clemson University, Clemson, SC 29634, USA, pmmurra@clemson.edu

Sakib Mahmud Khan, S.M., Khan[¶]

Simulation and Modeling Principal, The MITRE Corporation, USA, sakibkhan@mitre.org

Venkat Krovi, V., Krovi

Professor, Department of Automotive Engineering, Clemson University, Greenville, SC 29607, USA, vkrovi@clemson.edu

Automated vehicle (AV) platooning has the potential to improve the safety, operational, and energy efficiency of surface transportation systems by limiting or eliminating human involvement in the driving tasks. The theoretical validity of the AV platooning strategies has been established and practical applications are being tested under real-world conditions. The emergence of sensors, communication, and control strategies has resulted in rapid and constant evolution of AV platooning strategies. In this paper, we review the state-of-the-art knowledge in AV longitudinal platoon formation using a five-component platooning framework, which includes vehicle model, information-receiving process, information flow topology, spacing policy, and controller and discuss the advantages and limitations of the components. Based on the discussion about existing strategies and associated limitations, potential future research directions are presented.

---

[§] DISTRIBUTION STATEMENT A. Approved for public release; distribution is unlimited. OPSEC# 9556.

[*] Corresponding author.

[¶] Approved for Public Release; Distribution Unlimited. Public Release Case Number 24-3964. The author's affiliation with The MITRE Corporation is provided for identification purposes only, and is not intended to convey or imply MITRE's concurrence with, or support for, the positions, opinions, or viewpoints expressed by the author.

## 1 INTRODUCTION

Platooning is a formation strategy in which a group of vehicles achieves and/or maintains a desired formation while navigating a route. Platooning offers considerable benefits in terms of improving roadway capacity and fuel economy [14, 131]. These benefits could be further enhanced when platooning is performed by a group of automated vehicles (AVs) compared to platooning of a group of human-driven vehicles due to the variable and longer reaction time of human drivers. Eliminating human involvement in the platoon operation, AV platooning could reduce human factor-related safety concerns and improve operational and energy efficiencies of roadway traffic movements.

AVs have in-built sensors, such as cameras, radio detection and ranging (radar) sensors, ultrasonic sensors, and light detection and ranging (lidar) sensors, that are used to receive surrounding information. In-vehicle computing units process the collected information to perceive the driving environment. and compute platooning parameters. For AV platooning, these suites of sensors and computing units also assist in collecting and processing motion information of neighboring vehicles, such as inter-vehicle distance, heading, and speed of the preceding and following vehicles. This information is then utilized by a platooning controller to compute the required control inputs for a subject vehicle to help achieve and/or maintain a desired formation. The efficacies of AV platooning are related to the formation geometry, operational algorithms, formation parameters, and the reliability of the sensors and communication systems used by the vehicles. Some of the formation parameters are either pre-defined (e.g., desired spacing between vehicles) and some others are captured in real-time (e.g., surrounding vehicle information).

AV platooning strategies include both longitudinal and lateral control strategies. The longitudinal control of AVs in a platoon maintains desired inter-vehicle spacings, which is crucial for avoiding collisions, ensuring smooth traffic flow, and enhancing the overall stability of the platoon, through precise management of acceleration and braking commands. An optimal inter-vehicle spacing not only enhances safety but also contributes to significant fuel efficiency gains due to reduced aerodynamic drag—a key advantage in vehicle platooning, particularly for heavy-duty vehicles such as trucks [119]. Longitudinal control strategies dominate the focus in AV platoon formation studies due to their significant impact on safety and fuel efficiency [219]. For common highway maneuvers, lateral control can be assumed to be relatively decoupled from longitudinal controls. Consequently, lateral control [2, 152, 208, 209], which is mainly responsible for keeping the vehicles within their designated lanes, has minimal influence on fuel efficiency and platoon stability. Thus, this review study focuses on longitudinal control strategy, which is the basis for AV platoon formation and most significantly impacts safety and fuel efficiency.

There are several other surveys on vehicle platooning strategies. For example, Kavathekar and Chen [90] reviewed the vehicle platoon-related research published over a period of 16 years, starting from 1994. The authors identified general categories to group the published research on platoons, which include obstacle sensing and avoidance, vehicle-to-vehicle communication, platoon string stability, trajectory planning, and platoon control strategies (i.e., longitudinal and lateral control). Later, Zheng *et al.* [235] formulated a four-component framework for automated platooning, which includes models of vehicles forming the platoon, information flow topology (among the platooned vehicles), platoon controller, and platoon formation geometry. However, the framework in [235] does not consider the impacts of latency in communication systems, vehicle heterogeneity in platoons, noise in the data captured by unreliable AV sensors or caused by malicious attackers, etc. Several studies focused on specific components of the four-component framework [7, 49, 170]. Feng *et al.* [49] expanded the four-component framework and considered the quality of communication (latency and packet loss) and perturbation caused by the platooning vehicles (i.e., sudden fluctuation in spacing, speed, or acceleration) as additional components. Soni and Hu [170] studied the platoon formation approaches in detail and categorized the formation approaches into three groups: (i) leader-follower group, where the followers track the leader, (ii) behavior-based group, where in a new environment, a coordinator monitors behaviors based on each AV data, and assigns tasks for each AV, and (iii) virtual structure group, in which virtual agents are assigned to follow a desired motion and real AVs are operated to follow the virtual agents' motion. Another survey conducted by Wang *et al.* [193] reviewed studies related to cooperative longitudinal motion control of several connected and automated vehicles (CAVs). The authors in [193] divided the CAV cooperative longitudinal motion control system into five modules as follows: (i) communication, (ii) localization, (iii) perception, (iv) planning, and (v) control. Moreover, Wang *et al.* [193] presented five transportation applications that could benefit from cooperative longitudinal motion control systems, namely, (i) cooperative adaptive cruise control (CACC), (ii) cooperative merging on highways, (iii) speed harmonization on highways, (iv) eco-driving at signalized intersections, and (v) coordination at unsignalized intersections. The authors in [193] identified the heterogeneity of vehicles, the communication issues, and the string stability, as the primary control issues faced by CAV cooperative longitudinal motion control strategies.

However, there still remains the need for a survey that accumulates, synthesizes, and provides a comparative presentation of the existing AV longitudinal platoon formation control strategies, and highlights the future scopes for research in different longitudinal platooning aspects, such as vehicle modeling, sensing, communications, information flow topologies, spacing policies, and controllers. The objective of this paper is to synthesize literature on AV longitudinal platooning strategies, identify challenges, and provide future research directions for AV longitudinal platooning. In section 2 of this paper, the contribution of this study is discussed. Section 3 presents a review of the existing studies based on a five-component platooning framework. In section 4, we discuss potential future research directions, and section 5 presents the summary and conclusions.

## 2 CONTRIBUTION

In this paper, we perform a comprehensive review of the existing AV longitudinal platoon formation studies. Research on AV platooning strategies has been ongoing for many years, leading to the development of numerous methods. However, a comprehensive survey addressing both theoretical and practical perspectives—while articulating current knowledge, identifying limitations and open challenges, and outlining paths to move forward—has been lacking. This study aims to fill that gap in the literature. In this section, we highlight the contributions of this survey.

First of all, there are several studies in the literature that performed reviews of the existing AV platoon formation strategies [90, 49, 170, 7, 102, 37, 103, 193, 85, 159, 135]. Table 1 presents a comparison of the scope of our survey with the existing platooning-related surveys, which shows that this survey provides a comprehensive overview of different aspects of AV platooning along with future scopes. Rather than presenting a general overview, some of the existing surveys primarily focus on specific aspects of AV platooning, for example, string stability [90], sensing and wireless communication [37, 159], human factors and driver characteristics [37], transitional maneuvers like merging, splitting, and lane changing [7], and distributed controllers [103], whereas [133] focuses on the alignment of clustering and vehicle platooning approaches. The other surveys, such as [85, 102, 170, 193], present a more general overview, rather than focusing on a few specific aspects, of AV platooning; however, among them, the latest overview-type survey [190] was published in 2019, after which a substantial amount of AV longitudinal platoon formation studies have been published in the last five years. Therefore, there remains a need for a comprehensive survey that presents an overview of the existing related studies from theoretical and practical perspectives, including the latest ones published in recent years, which is covered in our survey.

Second, utilizing a five-component AV longitudinal platoon formation framework, this survey categorizes a vast body of existing studies, including the latest related studies, and presents each variation under an operational component of the framework with a set of common mathematical representations. For example, the vehicle model component in our five-component framework includes three variations: (i) first-order model, (ii) second-order model, and (iii) third-order model; each of which has been presented with a set of common mathematical representations in section 3.1 (i.e., Vehicle Model) of this survey. This comprehensive categorization of existing studies, including a common mathematical representation for each variation, enabled us to identify their common as well as specific strengths and weaknesses.

Third, we present several open challenges to AV longitudinal platoon formation in section 4 (i.e., Future Research Directions), which need to be addressed in future research to bring wide-scale adoption of AV platooning strategies to reality. These open challenges, presented for each of the five components of the AV longitudinal platoon formation framework used in this study, evolved from a detailed analysis of existing shortcomings of the related strategies which are covered in section 3 (i.e., AV Longitudinal Platoon Formation Framework).

Table 1: Comparison of the scope of existing surveys with this survey

| Survey Source | Survey Year | Vehicle Model | Information-receiving process | Information Flow Topology | Spacing Policy | Controller | Future Research Directions |
|---|---|---|---|---|---|---|---|
| [90] | 2011 | NP | P | NP | NP | P | P |
| [102] | 2015 | P | NP | P | P | P | NP |
| [85] | 2016 | NP | P | P | PP | NP | P |
| [37] | 2016 | NP | P | NP | PP | P | P |
| [100] | 2017 | P | PP | P | P | P | P |
| [170] | 2018 | PP | NP | P | NP | PP | P |
| [45] | 2019 | PP | NP | P | NP | PP | P |
| [159] | 2019 | NP | P | P | PP | P | P |
| [193] | 2019 | NP | PP | P | PP | P | P |
| [7] | 2021 | P | PP | NP | PP | P | P |
| [135] | 2024 | NP | P | PP | PP | P | PP |
| **Our Survey** | **2025** | **P** | **P** | **P** | **P** | **P** | **P** |

*P: present; PP: partially present; NP: not present*

## 3 AV LONGITUDINAL PLATOON FORMATION FRAMEWORK

The primary objective of an AV longitudinal platoon formation control strategy is to maintain stable platoon formation. This formation needs to follow a desired formation geometry, which, in the case of AV platooning, can be defined by an inter-vehicle spacing policy (e.g., constant spacing policy), as shown with a green arrow in Figure 1. To track this desired inter-vehicle spacing, a controller is needed for each AV in a platoon. The controller (indicated with the blue-colored boxes in Figure 1) of a subject AV in a platoon requires motion-related data (e.g., position and speed) of the subject AV and from the subject AV's neighboring vehicles. This data can be obtained via on-board sensors, such as inertial measurement units (IMUs), cameras, radars, lidars, and global positioning system (GPS) sensors, or via on-board communication devices, such as cellular vehicle-to-everything (C-V2X) direct communication devices (also indicated in Figure 1). Moreover, depending on the control objectives, the subject vehicle may require this data from some selected vehicles in the platoon only; for example, a subject AV may need the motion-related data only from its immediate predecessor and the platoon leader. Thus, an AV longitudinal platoon formation control strategy also needs to define an information flow topology that presents how these motion-related data will flow among the AVs in a platoon. A leader and predecessor-following topology is shown in Figure 1 with orange arrows as an example topology. Finally, an AV longitudinal platoon formation control strategy must define a controller that will provide control inputs (e.g., throttle and brake commands) to a subject AV that are essential for operating the AV. These control inputs are determined based on the current states of the subject AV and the current states of the neighboring vehicles of the subject AV as well. A vehicle model (indicated with red-colored boxes) is therefore needed to realize how the chosen control inputs will affect the subject AV or the platoon based on the current states of the platoon, which helps in determining the most suitable control actions at a given timestamp. Therefore, in this study, we review the existing AV longitudinal platoon formation studies based on a five-component framework, as presented in Figure 1. In this section, we discuss in detail how researchers have incorporated these components in their AV platooning studies.

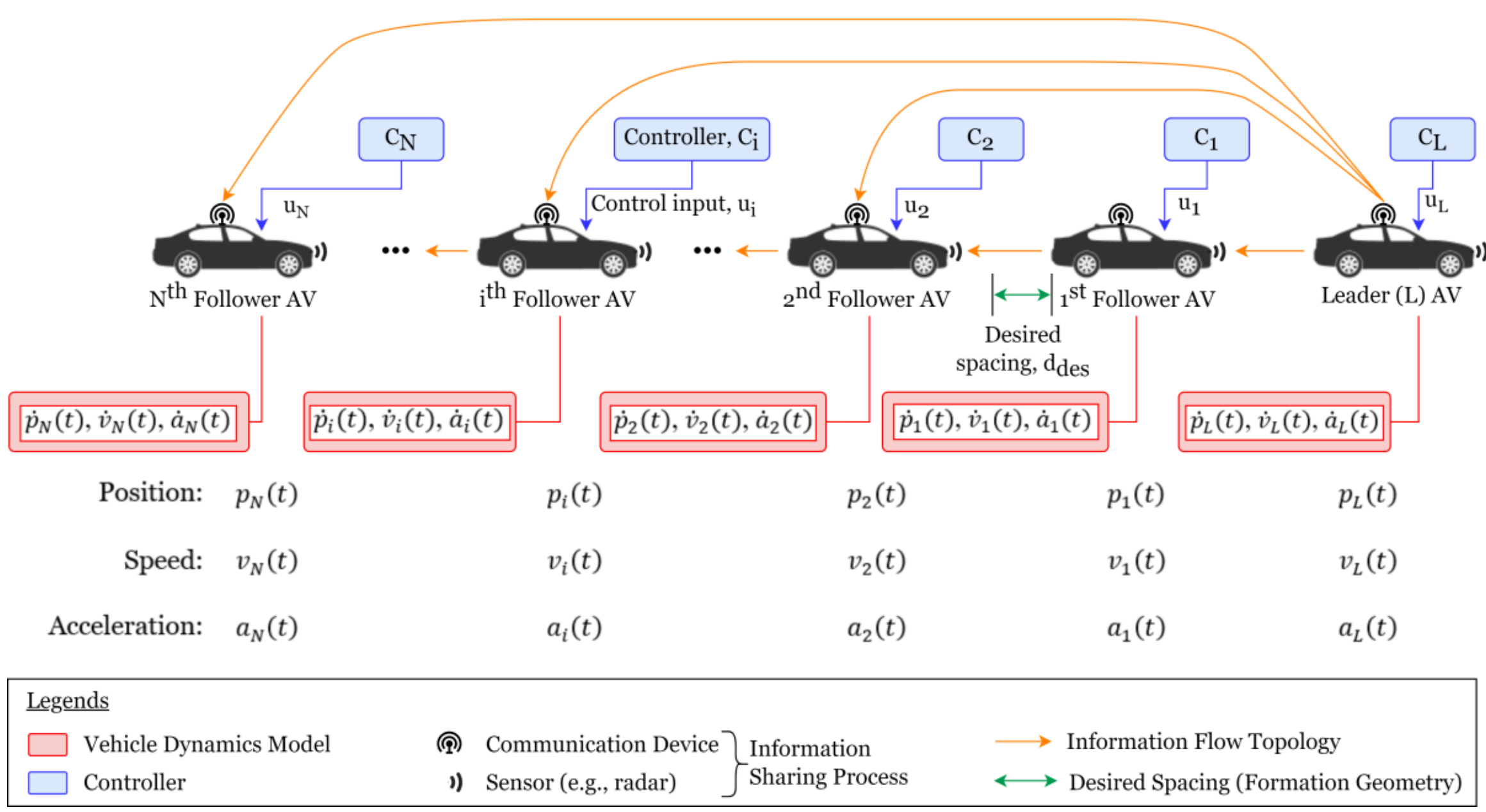

Figure 1: AV Platooning Framework.

### 3.1 Vehicle Model

Vehicle models play a significant role in designing the longitudinal control strategy for an AV platoon. The characteristics of vehicle models can vary from simple low-fidelity models to complex high-fidelity models, contingent on the designer's objectives (desired level of fidelity). Often, the modeling of vehicles in longitudinal, lateral, and vertical directions can be decoupled to simplify the analysis, as motion in each direction is governed by distinct effects (driving torque, tire slip angles, and road characteristics) and does not significantly impact behavior in other directions in normal conditions. The combined effects of these models need to be accounted for in complex driving conditions, such as braking/acceleration while cornering on a complex terrain. However, high-fidelity vehicle models can be highly nonlinear and computationally expensive for on-board processing. Therefore, researchers tend to utilize simplified or linearized models for vehicles in an AV platoon.

In Table 2, we present a list of different longitudinal vehicle models used in the existing AV platooning studies. Here, the longitudinal vehicle models are categorized based on the order of differential equations, with the vehicle's position being one of the dependent variable states. As discussed in [111], Some of the nominal models utilized in upper-level controllers in hierarchical controller design are double integrator models ($\ddot{p}(t) = u(t)$) and third-order models ($\tau\dddot{p}(t) + \ddot{p}(t) = \ddot{p}_{des}$), where, $p$ and $p_{des}$ denote the actual and the desired positions of an AV, respectively, and $u$ is the control input. These representative models often do not account for nonlinearities in AV powertrain systems (e.g., engine, drivetrain, and brake system) and aerodynamic drag (nonlinear dependence on velocity). In the literature, different studies considered the nonlinearity in longitudinal modeling to different extents, whereas others have utilized linear models to express longitudinal behaviors, which we discuss in details in this subsection. To present various types of longitudinal models used in the existing AV platooning literature, we categorize these models as follows: (i) first-order, (ii) second-order, and (iii) third-order models.

Table 2: Categories of existing AV platooning studies based on vehicle longitudinal models

| | |
|---|---|
| First-order model | [46, 116, 117, 145, 182, 220] |
| Second-order model | [10, 12, 15, 22, 25, 28, 29, 36, 38, 40, 45, 66–69, 73, 79, 86, 87, 89, 92, 93, 97, 99, 106–108, 110–113, 116, 144, 147, 155, 158, 166, 169, 174, 175, 178, 182, 183, 185–189, 192, 206, 207, 210, 218, 221, 223, 231] |
| Third-order model | [16, 23, 24, 26, 27, 42, 48, 55, 56, 58–65, 74, 94, 95, 101, 109, 124, 127, 128, 134, 136, 137, 141–143, 148–150, 152, 151, 156, 157, 163, 165, 171, 176, 179, 198–200, 203, 214, 225, 233–237, 239, 240, 242, 243] |

#### *3.1.1 First-order Model*

Single-integrator models are a class of first-order systems that consider the speed of a subject AV as the control input and its inertial position as the state [46, 116, 145, 182]. These models directly relate the rate of change in position of a subject AV along the longitudinal axis to the control input, i.e., speed. The general representation of the first-order longitudinal models in the literature can be written as follows,

$$\dot{p}_i = u_i + \omega_i \tag{1}$$

where, $u_i$ is the control input (i.e., speed, in this case) to the $i$-th follower AV in the platoon, $p_i$ is the position of the $i$-th follower AV, and $\omega_i$ is the external disturbance or noise associated with the $i$-th follower AV, respectively.

As observed in (1), the first-order models significantly simplify the longitudinal dynamics of the AVs in a platoon. This simplification, along with linearity, often helps convert the design of an AV longitudinal platoon formation optimal controller into a convex (quadratic) optimization problem and reduces computational complexity, as observed in [46, 116]. Authors in [46, 116] used simplified linear models in the formulation of optimal control problems to study the effects of increased communication links and communication with nearest neighbors with increasing platoon size. First-order models are deployed in distributed control algorithms and consensus algorithms, where the primary focus is achieving a set of predefined common goals for all the AVs in a platoon. Due to their minimal onboard computational cost, the first-order models serve as proof of concept for developing the control algorithms. Also, due to simplicity, first-order models can be useful to better understand how information flow topologies can affect AV longitudinal platooning performance [83, 129, 140, 201]. However, this simplification substantially deviates from the actual nonlinear longitudinal behavior of AVs in a platoon, causing controllers to suffer from issues such as string instability, i.e., increment in perturbations such as gap error throughout the AVs in a platoon.

### *3.1.2 Second-order Model*

Second-order models have been widely used in the existing AV longitudinal platoon formation studies. The second-order models that describe the dynamics of the vehicles in a platoon can be divided into two subcategories: (i) non-linear models and (ii) double integrator models. Non-linear models in the AV platooning literature (e.g., [22, 79, 99, 169, 218]) can be presented as follows,

$$\dot{p}_i = v_i \tag{2a}$$

$$\dot{v}_i = u_i - F_i + \omega_i \tag{2b}$$

where, $p_i$, $v_i$, and $u_i$ are the position, speed, and control input (e.g., acceleration or driving force) of the $i$-th follower AV in the platoon, $u_i$ is the control input to that AV, $\omega_i$ is the external disturbance or noise associated with the AV, and $F_i$ is the sum of the driving resistances for the AV, which can be written as,

$$F_i = F_r + F_a + F_g \tag{2c}$$

where, $F_r = c_{0,i} + c_{1,i} v_i$,

$F_a = \frac{1}{2}\rho C_{d,i} A_i v_i^2$, and

$F_g = g sin(\varphi)$

Here, $F_r$, $F_a$, and $F_g$ are the rolling resistance, the aerodynamic drag, and the resistance due to road grades, respectively; $c_{0,i}$ and $c_{1,i}$ are the coefficients of rolling resistance; $\rho$, $C_{d,i}$, and $A_i$ are air density, the coefficient of aerodynamic drag, and the cross-sectional projected area of the $i$-th AV, respectively; and $g$ and $\varphi$ are the gravitational constant and the road grade, respectively.

Feedback linearization is often used in studies to transform nonlinear systems into linear systems to ease designing a control law that could help achieve a desired input-output system behavior [171, 173]. Several studies in the literature utilized this technique to obtain a linear double-integrator model [10, 12, 15, 25, 28, 29, 38, 66–69, 86, 87, 89, 97, 106, 108, 110–113, 147, 158, 182, 189, 192, 221, 231], which can be written as,

$$\dot{p}_i = v_i \tag{3a}$$

$$\dot{v}_i = u_i + c_i v_i + \omega_i \tag{3b}$$

where, $c_i$ is the drag coefficient and $\omega_i$ is the disturbance or noise associated with the $i$-th follower AV. The other terms in (3) hold the same meanings as before. Some studies utilize feedback linearization to cancel out the effect of nonlinear aerodynamic drag, which further simplifies the model [12, 15, 25, 28, 29, 38, 66–69, 86, 87, 97, 106, 110–113, 158, 182, 189, 192, 221, 231].

The double integrator models describe the vehicles in the AV platoon where a control input directly controls the acceleration of the vehicle, and is given by [45, 73, 155, 178, 186–188],

$$\dot{p}_i = v_i \tag{4a}$$

$$\dot{v}_i = u_i = k_{d_1}(p_{i-1} - p_i - d_{des}) + k_{v_1}(v_{i-1} - v_i) \tag{4b}$$

$$\dot{v}_i = u_i = k_{d_1}(p_{i-1} - p_i - d_{des}) + k_{d_2}(p_i - p_{i+1} - d_{des}) + k_{v_1}(v_{i-1} - v_i) + k_{v_2}(v_i - v_{i+1}) \tag{4c}$$

Note that (4a) and (4b) are applicable for unidirectional flow of information, whereas (4a) and (4c) are applicable for bidirectional flow of information. Here, $k_{d_1}$ and $k_{d_2}$ are the control gains for the relative distances of the $i$-th follower AV with respect to the $(i-1)$-th and $(i+1)$-th follower AVs, respectively; $d_{des}$ is the desired inter-vehicular distance; and $k_{v_1}$ and $k_{v_2}$ are the control gains for the relative speeds of the $i$-th follower AV with respect to the $(i-1)$-th and $(i+1)$-th follower AVs, respectively. The other terms in (4) hold the same meanings as before. In (4c), when $k_{d_1} = k_{d_2}$ and $k_{v_1} = k_{v_2}$, i.e., motion-related data from both immediate neighboring AVs are weighted equally, then the corresponding control law is *symmetric* (e.g., [45, 73, 186–188]); otherwise, it is *asymmetric* (e.g., [155, 178]).

The double integrator models, in comparison to nonlinear models, are relatively simpler and do not capture sufficient dynamics of the vehicles in a platoon. They are used to achieve various objectives, such as coherence in large-scale networks [10], improvement of closed-loop stability margin [12], and stability and robustness of large platoons [68]. Furthermore, the double integrator models can be used to model platoon dynamics in the form of partial differential equations (PDEs) to capture the effects of disturbances and improve stability.

As observed from (2)-(4), second-order models assume that the acceleration of the AVs can be directly controlled and often do not take inertial or parasitic delays and lags into consideration, which is not practical. Note that parasitic delays and lags include (i) delays associated with the actuators and sensors in the AVs, for example, delays and lags associated with the powertrain system, and (ii) lags associated with filters used for sensors, such radars, pressure sensors, and wheel speed sensors [153, 154, 202, 203]. Neglecting these delays can cause instability in a real-world driving scenario [94, 103, 148, 171, 202]. Therefore, second-order models are also not practical in terms of replicating real-world driving conditions [203]. As a result, these models are often considered in only upper-level control models in a hierarchical control structure, where a lower-level controller maps the desired acceleration based on the upper-level control model to the required throttle input.

#### *3.1.3 Third-order Model*

As parasitic delays and lags are critical considerations for real-world implementation of AV longitudinal platoon formation strategies, many studies [15, 23, 24, 26, 27, 42, 48, 55, 56, 58–62, 64, 65, 74, 94, 95, 101, 109, 124, 127, 128, 134, 136, 137, 141–143, 148–152, 156, 157, 163, 165, 171, 176, 179, 198–200, 203, 214, 233–237, 239, 240, 242, 243] incorporate a third-order model by taking acceleration as an additional state variable to the second-order models. A general nonlinear third-order model to represent the vehicle longitudinal dynamics can be written as follows,

$$\dot{p}_i = v_i \tag{5a}$$

$$\dot{v}_i = a_i \tag{5b}$$

$$\dot{a}_i = f_i(v_i, a_i) + g_i(v_i)b_i + \omega_i \tag{5c}$$

where, $p_i$, $v_i$, and $a_i$ are the position, speed, and acceleration of the $i$-th follower AV in a platoon, respectively; $b_i$ is the control input to the engine; $\omega_i$ is the external disturbance associated with the AV; and $f_i$ and $g_i$ are nonlinear functions given by,

$$f_i(v_i, a_i) = -\frac{1}{\tau_i}\left[a_i + \frac{\rho C_{d,i} A_i}{2m_i} v_i^2 + \frac{d_{m,i}}{m_i}\right] - \frac{\rho C_{d,i} A_i}{m_i} v_i a_i \tag{5d}$$

$$g_i(v_i) = \frac{1}{m_i \tau_i} \tag{5e}$$

where, $m_i$ $C_{d,i}$, and $A_i$ are the mass, the coefficient of aerodynamic drag, and the windward area of the $i$-th follower AV, respectively; $\rho$ is air density; and $d_{m,i}$ and $\tau_i$ are the mechanical drag or resistance and engine time constant (also known as the time constant of the first-order inertial lag or "lumped" parasitic lag) of the $i$-th follower AV, respectively. A widely adopted approach in platooning studies is to transform the system in (5) through feedback linearization by taking,

$$b_i = m_i u_i(t - \Delta_i) + \frac{1}{2}\rho C_{d,i} A_i v_i^2 + d_{m,i} + \tau_i \rho C_{d,i} A_i v_i a_i \tag{6}$$

where, $\Delta_i$ is the "lumped" parasitic delay associated with the $i$-th follower AV, and $u_i(t - \Delta_i)$ is a new control input to the $i$-th follower AV from the feedback linearization. Substituting (6) into (5), the third-order vehicle longitudinal dynamics model can be rewritten in a linear form as,

$$\dot{p}_i = v_i \tag{7a}$$

$$\dot{v}_i = a_i \tag{7b}$$

$$\dot{a}_i = \frac{1}{\tau_i}[u_i(t - \Delta_i) - a_i] + \omega_i \tag{7c}$$

Some studies utilize this model for the upper-level controller in a hierarchical control architecture [56, 58]. The equations in (5)-(7) can also be written in terms of driving and braking torques (e.g., [42, 203, 235–237]). The authors in [13] control the driving/braking torque directly to improve the resiliency of the heterogeneous platoon to external attacks.

Another approach adopted by some studies [4, 5, 32, 52, 53, 224, 227] considers the change in speed and torque from the engine to the transmission system. A general representation of these nonlinear models is given by,

$$\dot{p}_i = v_i \tag{8a}$$

$$m_i \dot{v}_i = \frac{T_d}{r_w} - F_i \tag{8b}$$

here, $F_i = \frac{T_b}{r_w} + F_g + F_a + F_r$

$$T_d = \eta_T i_g i_0 T_t$$

$$T_t = k_{TC} T_p$$

$$T_p = T_e - J_e \dot{\omega}_e$$

$$T_e = \frac{1}{\tau_e s + 1} T_{es}$$

$$T_{es} = MAP(\omega_e, \alpha_{thr})$$

$$T_b = \frac{k_b}{\tau_b s + 1} P_{brk}$$

$$F_g = m_i g sin(\varphi)$$

$$F_a = \frac{1}{2}\rho C_{d,i} A_i (v_i + v_w)^2$$
$$F_r = m_i g f_r cos(\varphi)$$

where, $m_i$, $v_i$, $r_w$, and $T_d$ are the mass, speed, effective wheel radius, and driving torque of the $i$-th follower AV, respectively; $F_i$ is the sum of the driving resistances experienced by the $i$-th follower AV, which includes the gravitational resistance (denoted by $F_g$), the aerodynamic drag (denoted by $F_a$), the rolling resistance (denoted by $F_r$), and the brake force (denoted by $\frac{T_b}{r_w}$); $T_b$ is the braking torque; $\omega_t$ and $T_t$ are the turbine speed and torque of the torque converter, respectively; $i_g$ and $i_0$ are the transmission gear ratio and the final gear ratio, respectively; $\eta_T$ is the mechanical efficiency of driveline; $k_{TC}$ and $T_p$ are the torque ratio and pump torque of the torque converter, respectively; $T_e$ and $T_{es}$ are the actual and static engine torque of the engine, respectively; $\omega_e$ and $\tau_e$ are the engine speed and time constant, respectively; $J_e$ is the moment of inertia of the flywheel; $MAP(*,*)$ is a nonlinear tabular function of the engine torque characteristics; $\alpha_{thr}$ is the throttle angle, which can be considered as the control input; $k_b$, $\tau_b$, and $P_{brk}$ are the total braking gain of the wheels, the time constant of the braking system, and the braking pressure, respectively; $g$ and $\varphi$ are the gravitational constant and the road grade, respectively; $\rho$ is the air density; $C_{d,i}$ and $A_i$ are the coefficient of aerodynamics drag and windward area of the $i$-th follower AV, respectively; $v_w$ is the wind speed; and $f_r$ is the coefficient of rolling resistance. Figure 2 presents a schematic diagram of the powertrain system to supplement the nonlinear model presented in (8).

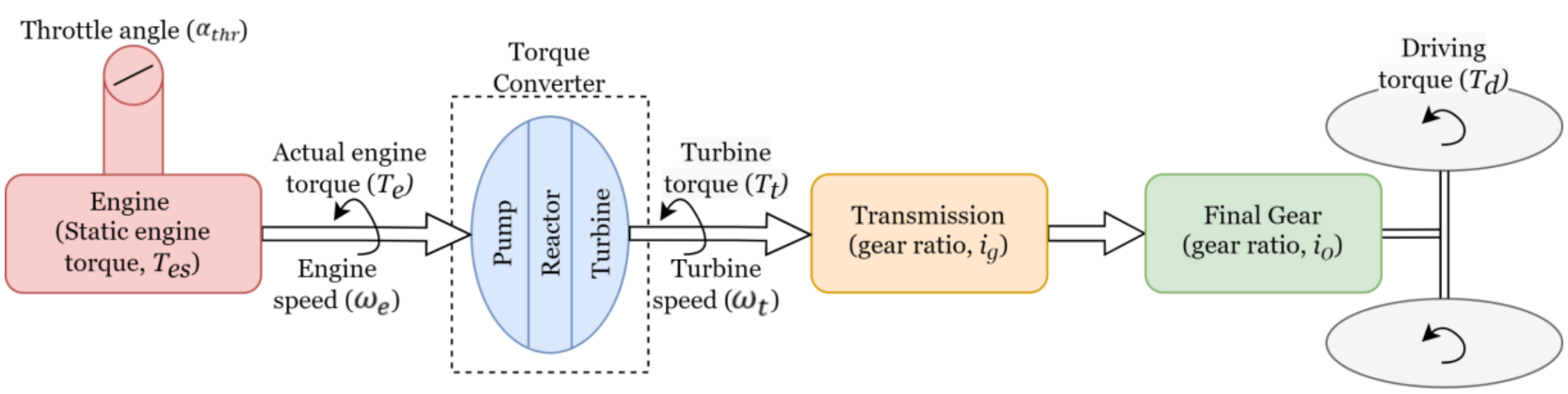

Figure 2: Mapping of the nonlinear model to the powertrain system.

### 3.2 Information-Receiving Process

To achieve and maintain a platoon formation, AVs are required to obtain neighboring vehicles' motion-related data, such as position, speed, and acceleration. This information can be obtained using (i) on-board sensors and/or (ii) on-board wireless communication devices. In this section, we discuss the information-receiving process assumed by the studies in the AV platooning literature (as listed in Table 3), along with the common challenges associated with it and the corresponding solutions discussed by the researchers.

Table 3: Categories of existing AV platooning studies based on information-receiving process

| | | | |
|---|---|---|---|
| Sensing | [4, 5, 15, 22, 32, 41, 42, 51, 55, 58, 59, 61, 68, 70, 71, 73, 86, 92–95, 99, 101, 104, 106, 108, 117, 128, 134, 143, 145, 161, 163, 166, 185, 187–189, 198, 206–209, 212, 214, 218, 221, 225, 235–237, 239, 242] | | |
| Communication | Packet dropout | No dropout | [4, 5, 15, 16, 22–27, 40, 42, 46, 48, 53, 56, 59, 60, 65, 74, 94, 95, 101, 106–112, 118, 124, 127, 133, 134, 136, 141–145, 147–152, 161, 163, 166, 171, 174, 178, 182, 185, 192, 198–200, 208, 210, 214, 218, 220–224, 229, 231, 233–237, 240, 242] |
| | | With dropout | [28, 29, 38, 52, 58, 79, 86, 87, 128, 137, 156, 158, 162, 165, 179, 189, 195, 206, 207, 209] |
| | Latency | Without latency | [4, 5, 22, 24, 38, 40, 42, 46, 48, 52, 59, 60, 65, 74, 94, 95, 101, 106–108, 110, 112, 127, 133, 137, 145, 147, 150–152, 161–163, 165, 166, 171, 174, 178, 182, 185, 192, 198–200, 210, 220, 223, 224, 231, 233–237] |
| | | With latency | [15, 16, 23, 25–29, 38, 53, 56, 58, 79, 86, 87, 109, 111, 118, 124, 128, 134, 136, 141–144, 148, 149, 156, 158, 179, 189, 195, 206–209, 214, 218, 221, 222, 229, 240, 242] |

### *3.2.1 Sensing*

AVs are typically equipped with various sensors that can be used to determine the relative distance and speed of neighboring vehicles. The selection of onboard sensors that are essential for longitudinal platooning could depend on different considerations, such as the set of information required for the platooning controllers, the level of measurement precision and sensor redundancy necessary for the selected controllers, the information flow topology being utilized, and the overall costs.

Among the wide variety of sensors considered for longitudinal platooning applications, radar, being a low to medium-cost sensor solution that can provide relative distance and speed information, is probably the most popular sensor [138, 139]. However, in some cases, radar-based distance and speed measurements may not provide precise enough data for longitudinal platooning control. Radar sensors also suffer from a limited field of view when compared to laser or lidar sensors. Thus, some studies considered laser radar and lidar sensors due to their higher precision data and broader field of view [1, 3, 197]. However, lidar or laser-based sensors are typically more expensive than short and long-range radars.

Sensor redundancy can be a crucial consideration for platooning strategies that rely on onboard sensors only unlike the strategies that consider wireless communication. Ogitsu *et al.* [139] presented to complement radar sensors with laser sensors to have one as a fallback system if the other one fails. Some studies present fusing data from multiple sensors for platooning operations. For example, Chakraborty *et al.* [21] presented fusing accelerometer and gyroscope data with global navigation satellite system (GNSS) data to supplement low-precision GNSS data for enhanced estimation of vehicle positions. Sensor fusion could provide robustness against noisy and erroneous measurements [21] and resilience against sensor attacks [126, 216]. While AV longitudinal platoon formation strategies with unidirectional information flow only need forward-facing sensors as a subject vehicle requires information regarding preceding vehicles only, strategies with bidirectional information flow often consider both forward- and rear-facing sensors as they require information from both preceding and follower vehicles [155, 187, 188].

Studies in the AV longitudinal platoon formation literature assumed different distributions, such as Gaussian and uniform distributions, to model the noise in sensor measurements [68, 86, 185]. A common approach to obtain state estimation via sensors is to utilize sensor fusion of multiple on-board sensors. In sensor fusion, filtering algorithms, such as Kalman filter, unscented Kalman filter, extended Kalman filter, and particle filter, are often used by a subject AV to estimate the states (e.g., position, speed) of the neighboring AVs in a platoon. Although there are various filtering techniques, in general, the most common filtering algorithms involve two steps: (i) a prediction step that predicts the current states of an AV based on its states information from the previous timestamp and (ii) a measurement step that compares the predicted states to the measured states [94].

Another issue associated with the data measured by on-board sensors is latency. A common approach to account for the latency associated with sensor measurements is to include a bounded latency term in the feedback controller. For example, Xu *et al.* [206, 208] included an equivalent latency into their control law to quantify the discretization in sensor data. It was shown in [75] that the latencies associated with discrete sensor data are bounded by the minimum sampling frequency of a sensor, which was adopted in [206, 208]. However, latency is a bigger concern in wireless communication than in sensing. Thus, most studies in literature tend to focus on latency associated with communication, which is discussed in the next subsection.

#### *3.2.2 Communication*

If AVs in a platoon are equipped with wireless communication devices, such as cellular vehicle-to-everything (C-V2X) devices, then the motion-related data can be directly sent to the other AVs in the platoon within the sender AV's communication range. As listed in Table 3, many studies in the literature assume perfect vehicle-to-vehicle (V2V) communication, i.e., no latency or packet dropout associated with V2V communication, among the platoon AVs. However, this assumption does not hold well in real-world driving conditions due to network-induced issues, such as transmission latencies, bandwidth limitations, and multiple connected AVs trying to share their information over the same network. Besides, on-board communication devices have limited communication range, which can be affected by vehicles' locations, speeds, and obstacles and cause loss of packets [18, 72]. Therefore, it is critical to consider issues, such as latency and packet loss, associated with V2V communication while designing an AV longitudinal platoon formation strategy [204, 205].

Studies in the AV longitudinal platoon formation literature often adopted some common assumptions to explore the effect of communication latency and packet dropout and to incorporate them into controller design as follows: (i) communication latencies are typically assumed to be bounded and detectable, and (ii) it is assumed that each AV that is receiving some information via wireless communication can estimate the latency incurred during transmission using the timestamp on the received data [15, 16, 26, 28, 29, 38, 58, 86, 87, 124, 158, 189]. A common practice for approximation-based latency estimation is to adopt the third-order Padé approximation, as shown in [85].

Researchers presented various strategies to account for communication imperfections, such as communication latency and packet dropout, in their developed control strategies. For example, Hao *et al.* [11] considered the aggregate communication latency as a bounded stochastic piecewise function in their AV platoon control law. Integration of motion-related data obtained via onboard sensors and wireless communication into the control law is utilized in some studies, for example, [15, 144, 158]. Chen *et al.* [29] designed a consensus algorithm that utilizes outdated motion-related data (e.g., position and speed) from the AVs in a platoon for homogeneous time-varying delay compensation. Wang *et al.* [189] considered switching the communication topology and the spacing policy if there is a communication failure. For example, while using a leader and predecessor-following (LPF) topology, if there occurs a communication failure with the leader, then the 1st follower AV can be assumed to be the leader by the rest of the AVs in the platoon, and the desired spacing can be increased to ensure safety. The authors in [189] included packet dropout as a latency of the data sampling period, which was also adopted by Guo and Yue [58]. The authors in [58] divided the motion-related data into two parts: information related to the preceding AV is obtained via sensors, whereas information related to the leader AV is obtained via wireless communication. Then, they introduced a symmetric logarithmic quantizer that maps the communicated information to different quantization levels to reduce the effect of delayed information. Öncü *et al.* [143, 144] introduced a sampled-data protocol-based network control system to account for communication delay. On the other hand, Hu *et al.* [79] assumed LPF communication topology with limited-range and low-volume information receiving to reduce the effect of communication latency and packet dropout. Sheikholeslam and Desoer [165] designed a control law to reduce the impact of communication loss with the leader AV on platoon performance. The control law minimizes the deviation of the 1st follower AV from its desired position when there is a change in the leader AV's speed from its steady state value. Naus *et al.* [136] used a system without any communication latency to design an estimator to estimate the acceleration of the preceding AV and later used that estimator in case communication with the preceding AV is interrupted. Chehardoli and Homaeinezhad [27] adopted a method introduced by Ergenc *et al.* [44], known as the cluster treatment characteristic root (CTCR) method, which determines the region of stability with respect to parasitic and communication latencies.

### 3.3 Information Flow Topology

Aside from the technology being utilized to send and/or receive the motion-related data, information flow topology varies largely depending on the information requirement set by the AV platoon controllers. Based on the existing AV longitudinal platoon formation studies, information flow topology can be broadly categorized as unidirectional, bidirectional, directed, and undirected flow of information (as listed in Table 4).

Table 4: Categories of existing AV platooning studies based on information flow topology

<table>
<tr><td rowspan="5">Unidirectional flow of information</td><td>Leader-following (LF)</td><td>[195, 223]</td></tr>
<tr><td>Predecessor-following (PF)</td><td>[4, 5, 22, 25, 41, 45, 48, 50–52, 64, 65, 68, 71, 74, 79, 92–96, 104, 112, 113, 118, 128, 134, 136, 137, 141–144, 148–150, 157, 158, 161, 164, 169, 176, 182, 183, 195, 203, 212, 221, 223–225, 229, 233, 236, 237, 239, 240]</td></tr>
<tr><td>Two predecessor-following (TPF)</td><td>[22, 52, 74, 118, 142, 148, 176, 233, 236, 237]</td></tr>
<tr><td>Leader and predecessor-following (LPF)</td><td>[15, 16, 29, 40, 45, 56, 58–61, 79, 109, 118, 124, 127, 141, 142, 147, 150–152, 156, 158, 161–165, 171, 174–176, 179, 189, 198, 206–209, 222, 223, 233, 236, 237]</td></tr>
<tr><td>Leader and two predecessor-following (LPTF)</td><td>[53, 142, 179, 223, 233, 236, 237]</td></tr>
<tr><td rowspan="3">Bidirectional flow of information</td><td>Bidirectional (BD)</td><td>[11, 12, 25, 32, 45, 52, 55, 62, 66–70, 73, 97, 99, 100, 116, 118, 133, 141, 142, 155, 158, 161, 178, 179, 187, 188, 192, 199, 218, 227, 234–237, 243]</td></tr>
<tr><td>Bidirectional leader (BDL)</td><td>[22, 27, 60, 100, 101, 106, 118, 142, 179, 199, 235–237]</td></tr>
<tr><td>Bidirectional double nearest neighbors (BDNN)</td><td>[199, 242]</td></tr>
<tr><td>Directed flow of information</td><td colspan="2">[23–26, 38, 47, 69, 86, 87, 89, 107, 111, 156, 179, 185, 195, 214, 220, 233, 237]</td></tr>
<tr><td>Undirected flow of information</td><td colspan="2">[36, 42, 46, 108, 110, 117, 145, 166, 189, 200, 210, 234, 235, 237]</td></tr>
</table>

*3.3.1 Unidirectional Flow of Information*

When the AVs in a platoon only require downstream or preceding AVs' information, it is called a unidirectional or unilateral flow of information. Unidirectional topologies can be divided further into (i) Leader-following or LF topology (where each follower AV in the platoon receives motion-related data from the leader AV of the platoon only), (ii) predecessor-following or PF topology (where each follower AV in a platoon only requires motion-related data from its predecessor), (iii) two predecessor-following or TPF topology (where each follower AV in a platoon requires motion-related data from its predecessor and the predecessor of its predecessor), (iv) leader and predecessor following or LPF topology (where each AV in a platoon requires motion-related data from its predecessor and the leader AV of the platoon), and (v) leader and two predecessor-following or LTPF topology (where each AV in a platoon requires motion-related data from its predecessor, the predecessor of its predecessor, and the platoon leader). Figure 3 illustrates these topologies, where the arrows represent the flow of information from one AV in a platoon to another. Because in a unidirectional flow of information, the follower AVs in a platoon receive motion-related data from the AVs in the downstream direction, any deviations of an AV in the platoon from its desired behavior only affect the AVs in the upstream direction. Thus, such deviations propagate through and get absorbed by only the upstream AVs for a string stable AV platoon using unidirectional flow information.

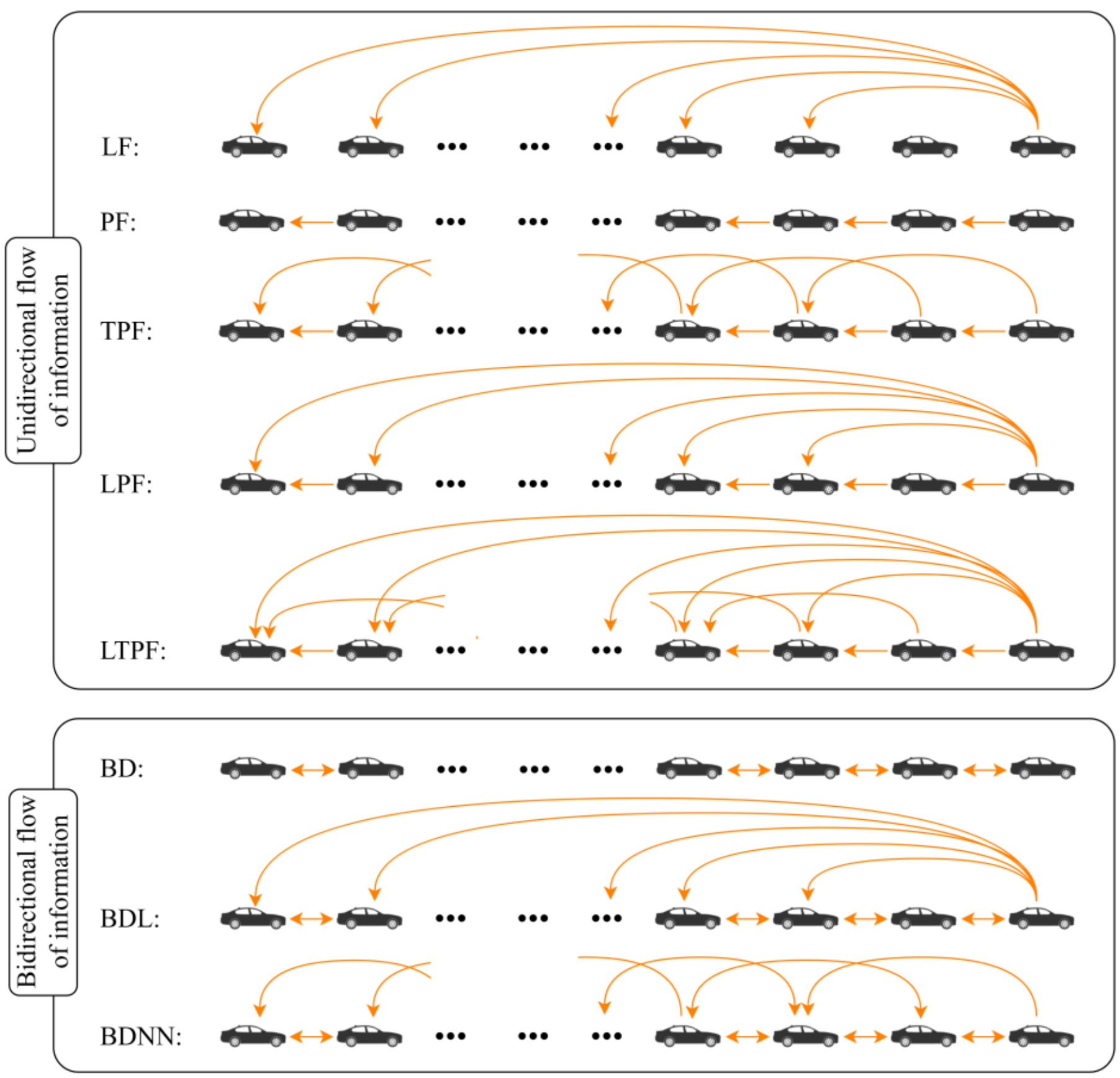

Figure 3: Different information flow topologies.

### *3.3.2 Bidirectional Flow of Information*

Some studies presented platoon controllers that require motion-related data from AVs in both the upstream and downstream directions. These are referred to as bidirectional (or bilateral) controllers. Studies that assumed a bidirectional flow of information used 1) bidirectional or BD topology (where each AV in a platoon requires motion-related data only from its immediate neighbors, i.e., its predecessor AV and follower AV), 2) bidirectional leader-following or BDL topology (where each AV in a platoon requires motion-related data from its immediate neighbors, i.e., its predecessor and follower AVs, as well as the leader AV), and 3) bidirectional double nearest neighbors or BDNN topology (where AV requires motion-related data from two of its immediate neighbors from each direction, i.e., two predecessors and two followers). Figure 3 illustrates the topologies under this category as well. Because in a bidirectional flow of information, the follower AVs in a platoon send/receive motion-related data to/from the AVs in both the upstream and downstream directions, any deviations of an AV in the platoon from its desired behavior affect the AVs both in the upstream and downstream directions. Thus, for a string stable AV platoon using a bidirectional flow of information, such deviations propagate through and get absorbed by the AVs in both the upstream and downstream directions, which helps in faster attenuation of disturbances than using unidirectional flow of information [73, 187].

### *3.3.3 Directed and Undirected Flow of Information*

Apart from unidirectional and bidirectional information flow topologies, some studies assume general information flow topologies. General topologies include directed and undirected information flow, where information flow is assumed not to be limited to nearest neighbors only (e.g., immediate predecessor and/or follower, and two predecessors and/or two followers) but beyond that, depending on factors such as the communication range. These topologies can be explained better using a graph theory approach. Assume that the information flow among the AVs in a platoon can be represented by a graph, $G = \{V, \varepsilon\}$, where $V = \{0,1,2, \dots, N\}$ is the set of nodes (i.e., AVs in the platoon), and $\varepsilon \subseteq V \times V$ is the set of edges among the nodes $V$ in a network. Then, the information flow topology for the AV platoon is considered as undirected if the graph $G$ is undirected, i.e., $j \epsilon \mathcal{N}_i \Leftrightarrow i \epsilon \mathcal{N}_j$, where $i, j = 1,2,3, \dots, N$ denote the follower AVs in the platoon, and $\mathcal{N}_i$ denotes the neighboring vehicles of the $i^{th}$ AV, otherwise, the information flow topology is said to be directed.

## 3.4 Spacing Policies

Spacing policies define the geometry of the AV longitudinal platoon formation. In general, the control objective of AV platoon controllers is to maintain a desired formation, i.e., maintaining a desired inter-vehicle spacing, while tracking the leader AV's speed,

$$\lim_{t \to \infty} \|v_i(t) - v_L(t)\| = 0 \tag{10a}$$

$$\lim_{t \to \infty} \|p_{i-1}(t) - p_i(t) - d_{i,des}(t)\| = 0 \tag{10b}$$

where, $p_i(t)$ and $v_i(t)$ are the position and speed of the $i$-th follower AV in a platoon at $t$, $v_L(t)$ is the speed of the leader AV of the platoon at $t$ timestamp, and $d_{i,des}(t)$ is the desired spacing of the $i$-th follower AV with its predecessor at $t$ timestamp, and we assume that the positions are measured from a point of reference that is located in the opposite direction of the platoon's heading. Spacing policy determines what $d_{i,des}(t)$ should be used for an AV platoon.

In general, there are three major spacing policies (as listed in Table 5): (i) constant distance headway (CDH) policy, (ii) constant time headway (CTH) policy, and (iii) variable time headway (VTH). In CDH policy, the AV longitudinal platoon formation controller tries to maintain constant inter-vehicle distance regardless of whatever speed the AVs are operating at, i.e.,

$$d_{i,des}(t) = d_{constant} \tag{11}$$

where, $d_{constant}$ denotes the constant inter-vehicle spacing or distance an AV platoon wants to maintain. In CTH policy, the AVs in a platoon want to maintain a constant time headway between each two AVs. Therefore, the desired spacing or distance may vary based on the operating speed of a particular AV. In CTH policy,

$$d_{i,des}(t) = d_0 + v_i(t) * T_h \tag{12}$$

where, $d_0$ is the standstill intervehicle spacing, $v_i$ is the speed of the $i^{th}$ follower AV, which is trying to maintain a desired gap (following the CTH policy) with its predecessor, and $T_h$ is a constant desired time headway. On the other hand, VTH is a nonlinear spacing policy, where the desired spacing for a follower AV in a platoon is a nonlinear function of that AV's speed, i.e.,

$$d_{i,des}(t) = d_0 + f\big(v_i(t)\big) \tag{13}$$

where, $f\big(v_i(t)\big)$ is a nonlinear function of $v_i(t)$. Note that the CDH policy can achieve high traffic throughput, whereas the CTH might have a lower traffic throughput than the CDH policy but better mimics human driving behaviors. On the other hand, the VTH policy has the potential to achieve high traffic throughput as well as maintain flow stability [157, 239].

Table 5: Categories of existing AV platooning studies based on spacing policy

| | |
|---|---|
| Constant distance headway (CDH) | [10–12, 23, 25, 27, 36, 40, 42, 45–47, 52, 53, 55, 56, 58–61, 65–70, 74, 87, 89, 92, 93, 97, 99, 100, 107, 110, 111, 116–118, 124, 127, 133, 136, 142, 145–147, 150–152, 161–165, 171, 174–176, 178, 179, 185, 195, 198–200, 206, 208–210, 214, 218, 220–224, 231, 233–237, 242] |
| Constant time headway (CTH) | [4, 5, 15, 16, 22, 24, 26, 32, 38, 45, 48, 50, 51, 62, 64, 79, 94–96, 101, 104, 112, 113, 128, 133, 134, 136, 137, 141, 143, 144, 148, 149, 155, 156, 158, 183, 189, 192, 203, 212, 222, 224, 225, 227, 229, 240] |
| Variable time headway (VTH) | [29, 109, 157, 239, 243] |

### 3.5 Controller

The objective of the controller in AV platooning is to determine specific control input to a subject AV in a platoon to achieve and/or maintain some desired formation based on the vehicle modeling assumptions and chosen spacing policy utilizing the motion-related data from the subject AV itself and its neighbors defined by the chosen information flow topology. Therefore, the controller in an AV longitudinal platoon formation strategy is a component where all other components (e.g., vehicle model, information-receiving process, information flow topology, and spacing policy) in AV longitudinal platoon formation framework come together to determine the control input based on a specific control law. Table 6 lists various types of controllers adopted in the existing AV platooning studies. Discussion on all the controllers listed in Table 6 is out of scope for this paper. Therefore, in this subsection, we choose to discuss the four most popular controllers among them, namely, (i) Linear control, (ii) Sliding mode control (SMC), (iii) $\mathcal{H}_\infty$ control, and (iv) Model predictive control (MPC). In addition, we present the latest developments in AV longitudinal platoon formation control strategies at the end of this section.

For each of the above controllers, we also discuss different approaches taken by researchers to ensure platoon performance, for example, local stability (which has to be guaranteed for any AV platoon controller), string stability, and stability margin of the platoon. A platoon of AVs with linear time-invariant (LTI) dynamics is considered stable if all the eigenvalues of each AV's closed-loop dynamics have strictly negative real parts.. Local stability implies the stability of individual AVs in a platoon. On the other hand, the stability margin implies the absolute value of the system's least stable eigenvalue's real part. String stability is achieved for an AV platoon when it is guaranteed that any perturbations, such as sudden braking of a vehicle within or in front of the AV platoon, will be dampened or absorbed across the upstream vehicles (for unidirectional flow of information) or both the upstream and downstream vehicles (for bidirectional flow of information). Apart from the abovementioned metrics, there are some other metrics used in the AV platooning literature, such as jerk or passenger comfort, safety, fuel economy, convergence, and traffic throughput or efficiency.

Table 6: Categories of existing AV platooning studies based on controller

| | |
|---|---|
| Linear control | [10–12, 15, 16, 23, 25–27, 32, 36, 38, 42, 45, 47, 55, 56, 66–70, 73, 86, 87, 92–94, 96, 97, 100, 107, 108, 133, 134, 136, 137, 142–145, 147, 149, 150, 155–158, 161, 162, 164, 165, 176, 178, 185, 187, 188, 195, 210, 214, 220, 221, 227, 235–237] |
| Sliding mode control (SMC) | [51, 52, 59, 60, 64, 65, 99, 101, 109, 112, 124, 146, 151, 152, 169, 175, 183, 192, 199, 203, 239, 243] |
| $\mathcal{H}_\infty$ control | [53, 58, 61, 148, 163, 206–209, 234, 240] |
| Model predictive control (MPC) | [40, 41, 48, 62, 79, 95, 104, 127, 179, 189, 198, 222–224, 231, 233] |
| Learning-based controllers | [2, 35, 39, 62, 63, 74, 81, 84, 123, 160, 181, 191, 194, 213, 228] |
| Other controllers | Different other optimal controllers: [5, 46, 89, 113, 116, 128, 171, 218, 224]<br>Nonlinear consensus: [29, 106, 110, 111]<br>Backstepping: [182, 242]<br>Passivity: [97]<br>Impulsive: [200]<br>Flatness-based: [50] |

### *3.5.1 Linear Control*

Linear control is the most widely utilized method among the existing AV longitudinal platoon formation studies (as listed in Table 6). These control methods gained substantial popularity due to their advantages in theoretical analysis, numerical analysis, and real-world deployment through hardware implementation compared to other types of controllers [236]. To provide a general representation of the linear controllers, we consider a third-order vehicle longitudinal model. Then, the tracking error for a desired spacing policy can be written as,

$$\boldsymbol{e}_{ij}(t) = \boldsymbol{x}_i(t) - \boldsymbol{x}_j(t) - \boldsymbol{d}_i(t) \tag{14}$$

$$\text{here, } \boldsymbol{x}_i(t) = [p_i(t), v_i(t), a_i(t)]^T$$

$$\boldsymbol{d}_i(t) = \left[d_{i,des}(t), 0,0\right]^T$$

where, $\boldsymbol{e}_{ij}(t)$ is the tracking error of the $i$-th AV with respect to the $j$-th AV in the platoon at time $t$; $j \epsilon \mathcal{N}_i$ with $\mathcal{N}_i$ denoting the set of the neighbors of the $i$-th follower AV; and the other symbols hold the same meaning as before. Then, the control law can be expressed as,

$$u_i(t) = -\sum_{j\epsilon\mathcal{N}_i} \boldsymbol{k}_{ij}\boldsymbol{e}_{ij}(t) \tag{15}$$

$$\text{here, } \boldsymbol{k}_{ij} = \left[k_{ij,p}, k_{ij,v}, k_{ij,a}\right]$$

where, $k_{ij,p}$, $k_{ij,v}$, and $k_{ij,a}$ are the control gains associated with position, speed, and acceleration. Although linear controls are popular among the platooning strategies presented in the literature, they often struggle with ensuring string stability and handling the nonlinearities and constraints involved in realistic driving conditions [193].

Studies that adopted a linear control considered different metrics to evaluate the platoon performance. First, local (asymptotic) or internal stability must be guaranteed for any AV longitudinal platooning strategies. For linear controls, the local stability of an AV platoon largely depends on the adopted information flow topology [237]. For instance, Lestas and Vinnicombe [72] provided a closed-loop stability theorem for establishing stability thresholds for various information flow topologies, such as PF, TPF, LPF, LPTF, BD, and BDL topologies. Second, different platoon control strategies based on linear consensus controllers have been developed to ensure string stability of an AV platoon, i.e., to ensure that any disturbances (e.g., errors in desired distance or speed tracking) imposed by a vehicle are dampened over the following vehicles in that platoon. For example, in [92, 93], Khaitr and Davison showed that while using identical controllers, it is impossible to ensure string stability for a homogeneous platoon of AVs under the CDH policy. This problem can be solved by using non-identical linear controllers, as presented in [93, 100, 164]. Another approach for solving the issue of string instability is to introduce the CTH policy as seen in [96, 173]. Stability margin has been adopted by several studies in the literature that incorporated linear controllers for AV platooning [12, 38, 42, 47, 66–69, 149, 185, 235, 236]. Barooah *et al.* [12] showed that the stability margin for an AV platoon is dependent on the size of the AV platoon as well as the information flow topology being considered. Using an asymmetric distributed linear control (as presented in [67]), the stability margin can be bounded away from zero, making it independent of the size of the AV platoon.

### *3.5.2 Sliding Model Control (SMC)*

Sliding mode control (SMC) is a variable structure nonlinear control strategy that utilizes a discontinuous control input to force a system to slide along or, in reality, "chatter" about some prescribed sliding surface after the system trajectory merges with the sliding surface in finite time. Different studies adopted different approaches for defining the sliding surface. For example, among the initial studies on SMC-based AV longitudinal platoon formation controllers, Rajamani *et al.* [151, 152] adopted the sliding surface design approach from [168] given by,

$$S_i = \dot{e}_i + \frac{\omega_n}{\zeta + \sqrt{\zeta^2 - 1}} \frac{e_i}{1 - C_1} + \frac{C_1}{1 - C_1}(v_i - v_L) \tag{16a}$$

$$\dot{S}_i + \lambda S_i = 0 \tag{16b}$$

$$\text{here, } \lambda = \omega_n\left(\zeta + \sqrt{\zeta^2 - 1}\right)$$

where, $S_i$ is the sliding surface; $\omega_n$ is a control gain; $\zeta$ is the damping ratio, which indicates how rapidly an oscillation decays over time and is set to 1 for critical damping; $C_1$ is a weight term; $e_i$ and $\dot{e}_i$ are tracking errors with respect to desired spacing and speed, respectively; and $v_i$ and $v_L$ denote the speed of the $i$-th AV and the leader AV, respectively.

Among the recent studies, Gao *et al.* [52] defined the sliding surface as follows,

$$S_i = a_i + \boldsymbol{K} \sum_{j \in \mathcal{N}_i} \begin{bmatrix} e_{ij} \\ \dot{e}_{ij} \end{bmatrix} \tag{17a}$$

here, $e_{ij} = p_i - p_j - d_{ij,des}$

and the corresponding control law is determined based on,

$$\dot{S}_i + \gamma_i S_i = 0 \tag{17b}$$

where, $\boldsymbol{K} \in \boldsymbol{R}^2$ is the control gain; $e_{ij}$ and $d_{ij,des}$ is the spacing error and the desired spacing between the $i$-th and $j$-th AV in the platoon, respectively; $j \in \mathcal{N}_i$ denotes the set of neighbors of the $i$-th AV; $\gamma_i > 0$ is responsible for the convergence speed of $S_i$; and other symbols hold the same meanings as before.

Studies incorporating SMC-based AV longitudinal platoon formation strategies took various approaches to ensure and evaluate the string stability of the platoon. For example, Xiao and Gao [203] presented how to design the control gains for their SMC to ensure that the practical string stability of both homogeneous and heterogeneous AV platoons would be maintained under the CTH policy in realistic conditions, such as experiencing time delays and lags. In another study [101], Li and Guo presented a new spacing policy by incorporating a platoon reference speed received by all the AVs in the platoon that is provided by a virtual leader acting as the upper-level platoon planning layer. Guo and Li [60] presented two SMCs for LPF and BDL topology-based AV platoons under the CDH policy that can guarantee individual vehicle's bounded stability as well as the whole platoon's string stability (i.e., errors are not amplified over the vehicles in the platoon).

SMC has several advantages as an AV longitudinal platoon formation controller, such as (i) SMC can be designed so that it is robust to external disturbances and model uncertainties, for example, by combining a switching control part with an equivalent controller [8, 52, 132], (ii) the switching function in SMC can be chosen to modify the controller's dynamic behavior [7], and (iii) SMC can directly specify platoon performance [101, 120]. On the other hand, SMC's main disadvantages are the "chattering" problem and not being able to handle constraints on the control input explicitly.

### 3.5.3 $\mathcal{H}_\infty$ control

H-infinity ($\mathcal{H}_\infty$) is a robust control design method that has been adopted by many studies. An AV platooning strategy must ensure that the platoon controller can handle uncertainties, such as a mismatch between the actual vehicle dynamics and the considered vehicle dynamic model, as well as the external disturbances caused by real-world driving conditions. In recent years, studies have shown how an $\mathcal{H}_\infty$ control design can be utilized for an AV platoon to handle such model uncertainties and external disturbances as well as sensor and communication delays while satisfying asymptotic and string stability criteria [58, 206–208]. To formulate a general representation of $\mathcal{H}_\infty$ control for an AV platoon, we consider a homogeneous platoon of AVs operating using the PF topology under a CDH policy. If we assume a second-order vehicle longitudinal model as presented in (3) and define,

$$e_i = p_{i-1} - p_i - d_{i,des} \tag{18a}$$

$$\dot{e}_i = v_{i-1} - v_i \tag{18b}$$

where, $e_i$ and $\dot{e}_i$ are the position and speed tracking error of the $i$-th AV in a platoon with respect to its predecessor AV, respectively; and the other symbols hold the same meaning as before. Then, we can write combining (3) and (18),

$$\dot{e}_i = v_{i-1} - v_i \tag{19a}$$

$$\ddot{e}_i = a_{i-1} - u_i - c_i v_i - \omega_i \tag{19b}$$

Then, we can define a linear control law as,

$$u_i = k_1 e_i + k_2 \dot{e}_i \tag{20}$$

where, $k_1$ and $k_2$ are control gains for the position and speed tracking, respectively. Plugging $u_i$ from (20) into (19) and taking $\xi_i = [e_i, \dot{e}_i]^T$, we can write,

$$\dot{\xi}_i = A\xi_i + HW_i \tag{21a}$$

$$Z_i = C\xi_i \tag{21b}$$

here, $A = \begin{bmatrix} 0 & 1 \\ -k_1 & -k_2 \end{bmatrix}$, $H = \begin{bmatrix} 0 & 0 & 0 \\ 1 & -c_i & -1 \end{bmatrix}$, and

$$W_i = [a_{i-1}, v_i, \omega_i]^T$$

Then, we can represent the cost as follows,

$$J(W_i) = \int_0^\infty \left[Z_i^T(t)Z_i(t) - \gamma^2 W_i^T(t)W_i(t)\right]dt \tag{22}$$

where, $\gamma \geq 0$ is a constant. Then, the optimal $\mathcal{H}_\infty$ control action, $u_i^*$, is to be determined to optimize the cost function defined in (22) while satisfying some predefined control objectives (e.g., some platoon consensus, fuel usage, and stability) along with $J(W_i) < 0$, $\forall W_i(t) \neq 0$ such that $\int_0^\infty \|W_i(t)\| dt < \infty$.

Studies that adopted $\mathcal{H}_\infty$ controllers for AV platoons took various approaches to ensure the stability of the platoon. For example, in [148], Ploeg *et al.* synthesized an $\mathcal{H}_\infty$ controller for an AV platoon, where the string stability condition was satisfied explicitly utilizing a linear matrix inequality (LMI). Zhou *et al.* [240] provided conditions for "head-to-tail" string stability, i.e., a disturbance is not amplified for a heterogeneous AV platoon system starting from the first vehicle to the last vehicle. In another study [234], Zheng *et al.* determined the stability margin of their developed $\mathcal{H}_\infty$ controller using a graph theoretic approach along with the Routh-Hurwitz criterion and presented how it can work for a wide range of information flow topologies.

As shown above, the $\mathcal{H}_\infty$ controllers formulate control problems in the form of optimization problems, and the optimal control law is determined by solving an optimization problem. $\mathcal{H}_\infty$ is a robust control method as it minimizes the effects of undesired signals and helps with stabilization. However, the real-world adaptation of an $\mathcal{H}_\infty$ controller is challenging for AV platooning since this control method also cannot handle constraints explicitly, though the constraints can be formulated as part of the cost function as "soft" constraints.

### *3.5.4 Model Predictive Control (MPC)*

MPC is an optimal control technique that predicts the future states of a plant based on the current states information and chosen control inputs. MPC utilizes numerical optimization to find the optimal control inputs while handling nonlinearity and satisfying a given set of constraints. As a result, many studies in the AV longitudinal platoon formation literature have adopted centralized MPC-based controllers, for example, see [189, 231], and distributed MPC-based controllers, for example, see [40, 48, 79, 95, 127, 179, 198, 223, 224, 233], for AV longitudinal platoon formation. However, a centralized MPC is not suitable for AV platooning application as it requires information of the states of all the AVs in the platoon. To provide a general representation of the distributed MPC-based controllers, we consider the third-order vehicle longitudinal model. Then, the plant model and its output can be written in discrete time as,

$$x_i^p(k+1|t) = \tilde{A}x_i^p(k|t) + \tilde{B}u_i^p(k|t) \quad (23a)$$

$$y_i^p(k+1|t) = \tilde{C}x_i^p(k+1|t) \quad (23b)$$

here, $\tilde{A} = I_{3\times3} + A\Delta t$, $\tilde{B} = B\Delta t$, and $\tilde{C} = C$

where, $x_i^p$, $y_i^p$, and $u_i^p$ are predicted state, output, and control input vectors at time $t + k\Delta t$, respectively, where $\Delta t$ denotes the time step size; and $I_{3\times3}$ is a 3-by-3 identity matrix. Then, the open-loop optimal control action can be written as,

$$u_i^*(t) = \underset{u_i^p(0|t),u_i^p(1|t),\ldots,u_i^p(N_c-1|t)}{\operatorname{argmin}} J_i(y_i^p, u_i^p, y_i^a, y_{-i}^a) \quad (24a)$$

$$\text{here, } J_i = \int_0^{N_p} l_i(y_i^p, u_i^p, y_i^a, y_{-i}^a)dt + \eta_i\left(y_i^p(N_p|t)\right) \quad (24b)$$

where, $u_i^*(t)$ is the optimal control input to the $i$-th AV in an AV platoon at time $t$; $y_i^a$ and $y_{-i}^a$ are assumed state vectors of the $i$-th AV and the neighbors of the $i$-th AV, respectively; $N_c$ and $N_p$ are the control and prediction horizons, respectively; $l_i$ and $\eta_i$ are the costs associated with $y_i^p$ until the final prediction step and at the final prediction step, respectively. Equation (24) can be subjected to a set of constraints depending on requirements, for example, the lower and upper bounds of speed and acceleration. Therefore, MPC is the only controller that is not limited by control saturation problems (i.e., situations where a feedback controller reaches a physical limit) unlike the other discussed control methods in this section. However, the challenge with an MPC is to select the state and control horizons and, more importantly, the cost at the final step in a way so that feasible solutions can be obtained continuously. Although MPC-based controllers are capable of handling constraints and nonlinearity explicitly, they, especially nonlinear MPC-based controllers, often suffer from high computational costs, which require AVs to be equipped with advanced on-board computing devices.

Studies that adopted centralized or distributed MPC for controlling AV platoons took various approaches to ensure the stability and robustness of the platoon. For example, Wang *et al.* [189] analyzed the asymptotic stability and the string stability for their centralized MPC-based AV platooning strategy. The authors in [189] adopted the asymptotic stability conditions for MPC developed by Mayne *et al.* [130] to ensure the asymptotic stability of a platoon and analyzed the string stability using the $\mathcal{L}_2$ string stability, i.e., spacing error attenuation notion of string stability, conditions presented in [49]. Dunbar and Caveney [40] ensured string stability to their distributed MPC-based platoon control strategy for PF and LF topologies by including additional constraints based on $\mathcal{L}_2$ string stability.

MPC is advantageous in AV platooning since it can provide an optimal control law that can explicitly handle constraints on the states and the control input. However, standard MPC requires a model of the plant dynamics (i.e., individual vehicle dynamics or dynamics of the whole platoon) that represent the plant well as MPC exploits the model to predict the future states (e.g., position and speed) of the system and optimizes the control law based on it. Therefore, standard MPCs cannot handle modeling or prediction errors, which is likely to occur in a real-world setting. This issue can be addressed using other variants of MPC, such as robust MPC [20], adaptive MPC [226], and tube-based MPC [115]. However, these methods could also result in a high computational cost for real-time AV platooning operations.

#### 3.5.5 *Learning-based Controllers*

With the advent of artificial intelligence (AI), AV platooning researchers have investigated different ways to leverage AI to control a platoon of AVs. In this subsection, we review studies that utilized AI within their AV longitudinal platoon control framework. Given the broader adoption of reinforcement learning (RL)-based approaches among the existing AI-supported AV longitudinal platoon formation studies, we divide our discussion in this section into two subsections: RL-based controllers and Other AI-based controllers.

##### 3.5.5.1 *RL-based Controllers*

RL is an established machine learning approach where an agent learns to take actions while interacting with an environment and aiming to maximize a reward function [172]. Inspired by behavioral psychology, RL enables learning through trial and error by providing rewards or penalties for actions. RL can be realized as a Markov decision process with a tuple $(\boldsymbol{X}, \boldsymbol{A}, \boldsymbol{P}, \boldsymbol{R})$, where $\boldsymbol{X}$ represents a state space, $\boldsymbol{A}$ represents potential action space, $\boldsymbol{P}$ represents a probabilistic state transition model, and $\boldsymbol{R}$ represents a reward function [98]. At time $t$, given a set of states $x_t$, the agent takes an action $a_t$ and the environment transitions according to $\boldsymbol{P}$. At time $(t+1)$, the agent observes another set of states $x_{t+1}$ following its action $a_t$ in the previous timestep and receives a reward $r_t$ according to $\boldsymbol{R}$. This iterates while the agent learns a policy $\pi(s_t, a_t)$ through observations and actions while maximizing the accumulated rewards.

Researchers have utilized this RL approach in different ways in the AV longitudinal platoon formation domain [98]. Among the early works, Dai et al. [35] developed a fuzzy RL-based AV longitudinal controller integrating a Q estimator network with a fuzzy inference system. The Q estimator network estimates an optimal value function, and the fuzzy inference system determines the control actions based on the value function. The reward in [35] accounts only for the spacing between a subject vehicle and its preceding vehicle in an AV platoon. Many studies have considered multi-objective reward functions to enhance AV platoon performance and stability. For instance, Yang et al. [213] presented a deep deterministic policy gradient (DDPG)-based proportional integral derivative (PID) control strategy to control the longitudinal motion in an AV platoon. In this method, the weights or gains of the PID controller are tuned dynamically leveraging a deep reinforcement learning framework. The authors of [213] utilized a hierarchical control architecture. In this architecture, an upper-level controller determines the desired acceleration of a subject vehicle in an AV platoon given the state information of the subject vehicle, its preceding vehicle, and the leader vehicle of the platoon. A lower-level controller translates the desired acceleration into gas/brake pedal input commands. The upper-level controller leverages an actor-critic network to determine the weights of the PID controller that constitutes the upper-level control input, i.e., the desired acceleration. The authors of [213] introduced a reward function that accounts for collision risks, stability, and response time to help train the actor-critic network. Another study [123] integrated deep reinforcement learning with dynamic programming to determine real-time control actions in an AV platoon by embedding the DDPG algorithm into a finite-horizon value iteration framework. The reward function considered in [123] is designed to minimize gap and speed errors while reducing fuel consumption and jerk.

Unlike other types of controllers that require a vehicle model which is difficult to develop due to the system's nonlinear nature, a model-free RL-based controller does not require an effective vehicle model and is able to learn an optimal control policy by interacting with the environment without any systems model [172]. However, the primary disadvantage of such controllers is the time required to train the agent to take optimal actions. Some studies addressed this issue by integrating RL with supervised learning to speed up the training process [181, 228].

#### *3.5.5.2 Other AI-based Controllers*

In addition to RL-based controllers, the dynamic nature of AV platooning encouraged researchers in recent years to explore adaptive neural network-based approaches due to their online learning capability [194]. In [194], Wei *et al.* leveraged an adaptive neural network to estimate the unknown vehicle dynamics and introduced the tangent barrier Lyapunov function into the control design. Their approach was reported to better handle external disturbances and actuator faults while maintaining Lyapunov stability. In another study [81], Huang *et al.* utilized an adaptive neural network to estimate the nonlinearity in vehicle modeling, external disturbances, and acceleration of neighboring vehicles. These approaches reduce the burden of ensuring robust sensory measurements and/or communication, as required by many platoon control strategies.

Controllers developed with data-driven AI-based methods aim to mitigate the effects of uncertainties and disturbances that are often challenging to model theoretically. Some efforts partially equipped the hierarchical control framework employed in longitudinal control for CACC with data-driven strategies, such as in [215, 241], where the upper-level controller alone was designed with long short-term memory (LSTM) and radial basis function neural network (RBFNN) to predict the desired accelerations. However, there has been research on completely substituting the control framework with data-driven controllers in an end-to-end driving framework as well [39, 191]. In [39], the author presented a data-driven control strategy leveraging gradient descent methods to address the cooperative adaptive optimal output regulation problem in continuous-time linear multi-agent systems. Instead of relying on model information, the approach in [39] computes gradients using real-time data, enabling the development of a data-driven distributed adaptive controller through adaptive dynamic programming. In another AI-based end-to-end driving study [191], the authors trained a Gaussian Process model with human drivers' longitudinal driving behaviors. The approach captured an individual driver's car-following behaviors and generated a desired acceleration profile that aligned with their preferences. The authors of [191] combined the trained model with a predictive safety filter designed to prevent rear-end collisions, developing a personalized adaptive cruise control system. Although these studies demonstrated the potential of AI models to replace traditional AV platoon controllers, their high reliance on real-world data poses a challenge.

### *3.5.6 Latest Developments in Controllers*

In this subsection, we discuss some of the recent efforts towards AV longitudinal platoon formation control, particularly with respect to the methodologies and considerations employed for controller development. Our review of the recent literature suggests that, over the past years, using optimization-based approaches to design stable, centralized/decentralized longitudinal control methods for AVs and AV platoons [9, 57] have gained popularity due to their ability to incorporate additional desirable performance metrics. Researchers presented solutions to optimize fuel consumption within stable platooning applications in [6, 17, 77, 80, 122, 125]. In [125], the authors incorporate the effect of aerodynamic drag for predicting desired accelerations to maintain string stability with vehicles powered by hydrogen fuel cells, whereas, the authors of [6] considered the effect of uneven terrain on fuel efficiency in platoons. The cooperative longitudinal control solutions presented in [82, 122] are applicable to urban settings. The optimization-based controller developed in [80, 125], accounted for passenger comfort in the optimization problem. Investigations presented in [6, 80, 91] considered vehicles with heterogenous dynamics in the formulation. The authors of [91] further incorporated disturbances in the form of human driver interferences in the design of their controllers. Optimal methods for safe and stable platooning applications, such as joining a pre-existing string with minimal disruption and smooth merging, were presented in [30, 105], and considering mixed traffic scenarios in [117, 166]. Solutions that are robust to communication issues were presented in [82, 91, 121, 177, 190, 230], whereas controllers robust to modeling uncertainties and external disturbances were developed in [79, 238].

While recent years have seen a rise in the popularity of optimization-based methods, classical and Lyapunov-based control methods are still employed for controller development for platooning applications due to the ease of their analyses utilizing well-established theoretical approaches. However, unlike optimization-based methods which are capable of handling multiple control objectives, the control objectives of traditional controllers are limited. The research in [31] investigated the issues of consistent spacing within heterogenous platoons and presented a novel prediction-based spacing strategy followed by a control scheme that reformulates the string to an LF-based consensus problem that is inherently decentralized in nature. Yamaguchi and Yang [211] followed a similar approach and developed a novel spacing policy integrating the concepts of constant distance headway with constant time headway. The presented spacing policy in [211] was verified with decentralized robust controllers. The authors of [167] presented an easy-to-implement, linear feedback controller with a constant time delay that could mitigate the effects of dynamic obstacles (e.g., a non-connected vehicle) in a CACC platoon. The challenges of modeling and parametric uncertainties and external disturbances on the internal and string stability of a platoon were mitigated in [54, 76, 78] with the design of a cascaded control framework. In this framework, the upper-level controller consists of a state observer that provides an estimate of the deviation of the internal states and the lower level is a tracking controller to compensate for the real-world uncertainties. The tracking controller in [54] was developed employing an adaptive scheme that allows the controller to be applicable for platoons of any size and for platooning operations like merging and splitting. The lower-level controllers in [76] and [78] were designed utilizing output synchronizing theory and sliding mode control approach, respectively. The studies presented in [196, 232] also utilized sliding mode control to mitigate the effects of inaccurate velocity measurements and intervehicle communication delays to ensure internal and string stability, smooth flow of traffic and passenger comfort. Modeling and parametric uncertainties were managed via adaptive control strategies, such as in [217], which used Lyapunov's methods to estimate parameters to develop an adaptive control law, similar to the model reference control methods. Wang *et al.* [184] presented an event-triggered control scheme while a robust and adaptive $\mathcal{H}_\infty$ gain scheduling method was developed in [43]. Some research has also incorporated game theory to control vehicles in strings and ensure collision-free navigation [88] as well as complex platoon maneuvers like on-ramp merging in mixed-traffic scenarios [114].

## 4 FUTURE RESEARCH DIRECTIONS

AV platoon formation has been an important topic of research and development for the last few decades. Although numerous studies in this field have been conducted over the years, several gaps remain where future research is needed before full-scale AV platooning can be realized in the real world. In this section, we shed light on several gaps with potential future research directions.

### 4.1 Vehicle Dynamics

An appropriate vehicle dynamics model is critical to maintaining a stable AV longitudinal platoon formation since it helps determine a suitable control law by estimating the effect of the chosen control input on individual vehicles as well as the overall platoon. In the existing literature, low-fidelity models capture kinematic data and are used to develop control laws to improve platoon stability and traffic flow with various information flow topologies. Higher fidelity models capture the internal dynamics of the vehicle and improve stability, but they often come at the cost of increased computational burden on the onboard processing unit. In the literature, the effects of the effective mass (inertia due to rotational components) are often not considered and would present an interesting opportunity to identify its impact in improving stability, fuel efficiency and traffic flow. The impact of powertrain models from traditional combustion engines to modern electric and hybrid models in terms of stability and fuel efficiency needs to be carefully understood.

### 4.2 Information-Receiving Process

In AVs, many sensors are utilized for perception and navigation. Each sensor has its own objective, advantages, and limitations. An interesting future research direction for AV longitudinal platoon formation is to explore the effects of fusing multiple on-board sensors as well as data received via wireless communication to obtain robust states information of the neighboring vehicles. Sensor fusion could enable fault-tolerant information-receiving systems for AVs. Many sensor fusion techniques have been presented in the literature; however, their usefulness in AV platooning strategies has yet to be explored. Also, latency is a major concern for wireless communication-based information receiving. 5G's reliable and faster communication has the potential to solve this issue. Researchers need to put more focus on studying the effects of these state-of-the-art technologies on AV platoon formation through real-world testing.

### 4.3 Spacing Policy

Various spacing policies have been introduced for AV longitudinal platoon formation over the years, considering various factors, such as traffic throughput, safety, fuel economy, passenger comfort, and communication range. However, their effect on platoon stability for heterogeneous roadway traffic and switching communication topologies is a matter of ongoing research. While leaving a larger gap among the vehicles in a platoon might be better for safety and stability, larger gaps reduce traffic throughput and increase fuel usage due to higher aerodynamic drag. Therefore, an optimal spacing policy that can optimize the above factors while helping the stable operation of a platoon is required. Besides, merging and diverging are very important considerations for AV platooning that are missing in many of the existing studies. Any AV platooning strategy should account for merging and diverging coordination among the AVs in a platoon and design the controllers accordingly.

### 4.4 Controller

The controller is a key component of AV longitudinal platoon formation. Over the years, many longitudinal controllers have been developed for AV platooning, as discussed in section 3.5. However, they all share a common issue of relying on an appropriate vehicle model, which is very challenging in a real-world scenario where there are heterogeneous vehicles on the road. In addition, vehicle model considerations and vehicle-specific parameters, such as size, shape, and weight, could also affect the performance of the platoon controllers. Although some studies have explored these effects for specific use cases, a comprehensive analysis of them remains as a gap yet to be addressed. To better understand the impact of different vehicle models and the corresponding parameters on platoon controllers, further theoretical and experimental investigations are necessary. Very recently, researchers have started to focus on designing model-free AV platooning strategies, such as neural network (NN)-based AV platooning strategies [2, 63, 84, 160]. However, the biggest challenge associated with such strategies is the requirement of a huge amount of data for training the NNs effectively to perform platooning operations. While a part of this data can be collected through high-fidelity simulations, data from the real world is still much needed.

## 4.5 Real-world Testing and Evaluation

Most of the AV platoon formation studies found in the literature focus primarily on simulation-based validation and lack real-world testing and validation. A few studies that conducted real-world tests to evaluate the performance of an AV platoon considered a controlled environment such as a test track or a closed-down section of a public roadway [19]. There are several real-world concerns with AV longitudinal platoon formation studies relying entirely on simulation-based validations, as discussed below.

### *4.5.1 Computational Load and Delay*

In most simulation-based evaluations, AV longitudinal platoon controllers utilize high-performing computing resources and may not always replicate strict real-time operation, i.e., simulations can be slowed down, or computations can be carried out offline. However, real-world deployment requires controllers to compute optimal control actions within milliseconds at each time step, posing challenges for many simulation-validated AV longitudinal platoon controllers' real-world deployment [180]. To ensure real-world deployment requirements are met, future studies should consider evaluating the real-time performance of their controllers with respect to commercially available onboard and edge computers' processing capabilities.

### *4.5.2 Model and Reality*

Simulations often use popular or well-known vehicle models, which could be very similar to the models utilized by an under-evaluation AV longitudinal platoon controller. This could potentially overestimate a controller's real-world performance since real vehicle kinematics and dynamics are quite complex and involve uncertainties. AV longitudinal platoon controllers relying on vehicle models and validated through simulations may not have the capacity to handle real-world complexity.

### *4.5.3 Sensor and Communication*

Sensors are often modeled in simulations following an ideal scenario and may not capture noises, delays, or measurement uncertainties to the extent that is observed in the real world. Similar to sensors, wireless communication in the real world is almost always prone to unwanted delays and packet drops. A simulated environment may not replicate all of these conditions effectively.

### *4.5.4 Dynamic Environment*

Real-world deployment of AV platoon formation strategies necessitates scalability and robustness [33]. In a real-world setting, the environment is always changing and includes multifaceted complexities, such as a sudden change in traffic conditions, mixed and heterogeneous traffic, changing pavement, weather, and lighting conditions, and unanticipated driving behaviors – a ready-to-deploy AV longitudinal platoon controller must be able to handle these dynamic conditions that a simulated environment might lack.

Therefore, it is important to consider the gaps between a simulated environment and a realistic driving environment while designing and evaluating an AV longitudinal platoon controller. According to [34], one should consider the physics fidelity (i.e., the degree to which a simulation resembles the real-world operational domain) and the functional fidelity (i.e., the degree to which a simulation behaves like an operational equipment while responding to executed tasks) of a simulated environment to realize the similarities and gaps between the simulated environment and a real-world environment. Although there exist several domain-specific high-fidelity simulators, such as microscopic traffic simulators, network communications simulators, sensor simulators, and physics engines, there is a need for a single simulator that could cover all these aspects alone. A co-simulation of different domain-specific high-fidelity simulators, such as the CARMA everything-in-the-loop (XiL) co-simulation tool [244] developed by the Federal Highway Administration (FHWA), could potentially bypass such limitations. Although these tools can be effectively utilized to conduct proof-of-concept evaluations, the need for wide-scale real-world testing and validation remains indispensable.

## 5 SUMMARY AND CONCLUSIONS

In this paper, we reviewed different AV longitudinal platoon formation strategies developed by researchers over the last few decades. A five-component platooning framework was considered to study the existing AV longitudinal platoon formation studies: vehicle model, process for receiving information, information flow topology, spacing policy, and controller. We categorized the existing AV longitudinal platoon formation studies based on the approach taken by the studies regarding each component of this framework. For each category (e.g., for each vehicle longitudinal dynamic model and for each controller type), we reviewed the corresponding studies to present the category with mathematical representations and discussed the strengths and weaknesses. For each AV longitudinal platoon formation controller, we also presented some exemplary approaches taken by the corresponding studies to ensure and evaluate the robustness and stability of the platoon. Finally, we discussed some future research directions that can lead researchers in this field to conduct further research toward making wide-scale adoption of AV platooning feasible.

Based on the review of the existing studies, we found some open challenges that need to be addressed for developing an AV longitudinal platoon formation. An AV longitudinal platoon formation strategy needs to account for the nonlinearity in vehicle dynamics and the delays and lags in the powertrain system while not overburdening the on-board processing unit. It also needs to consider common issues with wireless communications, such as latencies and packet dropouts, and adapt to switching among the most suitable information flow topologies when needed. Stability and robustness to external disturbances are crucial for an AV longitudinal platoon formation controller to be implemented in the real world. While meeting all these requirements is challenging, there remain additional practical requirements, such as improving the roadway traffic throughput/capacity, improving safety, reducing aerodynamic drag, improving overall fuel efficiency, and considering passenger comfort.

## ACKNOWLEDGMENTS

This work was supported by Clemson University's Virtual Prototyping of Autonomy Enabled Ground Systems (VIPR-GS), under Cooperative Agreement W56HZV-21-2-0001 with the US Army DEVCOM Ground Vehicle Systems Center (GVSC).

## 6 HISTORY DATES